\documentclass[prl,aps,twocolumn,floatfix,preprintnumbers,superscriptaddress]{revtex4-2}

\usepackage{amsfonts}
\usepackage{mathtools}
\usepackage{lmodern}
\usepackage[T1]{fontenc}
\usepackage[utf8]{inputenc}
\usepackage{bbm}
\DeclareMathAlphabet{\mathbfi}{OML}{cmm}{b}{it}

\usepackage{hyperref}
\usepackage{xcolor}

\let\originalleft\left
\let\originalright\right
\renewcommand{\left}{\mathopen{}\mathclose\bgroup\originalleft}
\renewcommand{\right}{\aftergroup\egroup\originalright}

\renewcommand{\vec}[1]{{\ifnum9<1#1\mathbf{#1}\else\ifcat\noexpand#1\relax\boldsymbol{#1}\else\mathbfi{#1}\fi\fi}}
\newcommand{\mathe}{\mathrm{e}}
\newcommand{\mathi}{\mathrm{i}}
\newcommand{\total}{\mathop{}\!\mathrm{d}}

\newcommand{\eqend}[1]{\,#1}

\makeatletter
\newcommand{\expect}{\@ifstar{\@expects}{\@expecta}}
\newcommand{\@expecta}[1]{\left\langle{#1}\right\rangle}
\newcommand{\@expects}[2]{#1\langle{#2}#1\rangle}
\makeatother

\begin{document}

\preprint{MITP-23-042}

\title{Universal definition of the non-conformal trace anomaly}

\author{Renata Ferrero} 
\email{renata.ferrero@fau.de}
\affiliation{Institut f{\"u}r Physik (THEP), Johannes Gutenberg-Universit{\"a}t Mainz, Staudingerweg 7, 55128 Mainz, Germany}
\affiliation{Institute for Quantum Gravity, Friedrich-Alexander-Universit{\"a}t Erlangen-N{\"u}rnberg, Staudtstr. 7, 91058 Erlangen, Germany}

\author{Sebasti\'an A. Franchino-Vi\~nas}
\email{s.franchino-vinas@hzdr.de}
\affiliation{Helmholtz-Zentrum Dresden-Rossendorf, Bautzner Landstra{\ss}e 400, 01328 Dresden, Germany}
\affiliation{DIME, Universit\`a di Genova, Via all'Opera Pia 15, 16145 Genova, Italy}

\author{Markus B. Fr\"ob} 
\email{mfroeb@itp.uni-leipzig.de}
\affiliation{Institut f{\"u}r Theoretische Physik, Universit{\"a}t Leipzig, Br{\"u}derstra{\ss}e 16, 04103 Leipzig, Germany}

\author{William C. C. Lima} 
\email{williamcclima@gmail.com}
\affiliation{Institut f{\"u}r Theoretische Physik, Universit{\"a}t Leipzig, Br{\"u}derstra{\ss}e 16, 04103 Leipzig, Germany}

\date{February 15, 2024}

\begin{abstract}
We show that there exists a generalized, universal notion of the trace anomaly for theories which are not conformally invariant at the classical level. The definition is suitable for any regularization scheme and clearly states to what extent the classical equations of motion should be used, thus resolving existing controversies surrounding previous proposals. Additionally, we exhibit the link between our definition of the anomaly and the functional Jacobian arising from a Weyl transformation.

\par
\begin{center}
\textit{{In memoriam} of Fidel Arturo Schaposnik.}
\end{center}

\end{abstract}

\maketitle

\section{Introduction}

Anomalies in quantum field theory (QFT) arise whenever a symmetry of a classical action cannot be preserved after quantization~\cite{Bertlmann,Fujikawa:2004cx}. While anomalies in local symmetries may render a theory inconsistent, anomalies of global symmetries can result in physical effects of major significance. 

Indeed, already the first anomaly described, the chiral anomaly of Adler, Bell and Jackiw~\cite{Adler:1969gk,Bell:1969ts}, resolved the difference between the theoretical and the observed decay rate of the process $\pi^0 \to 2 \gamma$. During the last decades, the range of applications of anomalies has pervaded all the areas of physics in which quantum effects play a privileged role, such as condensed matter~\cite{Chernodub:2017jcp, Nissinen:2019kld, Chernodub:2019tsx}, astrophysics~\cite{Schutzhold:2002sg, Fujimoto:2022ohj}, effective field theories~\cite{Filoche:2022dxl,Cohen:2023hmq}, as well as more formal aspects in quantum field theory~\cite{Jensen:2018rxu,Coriano:2020ees,Boos:2020ccj, Fan:2022ukm,Barvinsky:2023aae} (including the celebrated connection between Hawking radiation and the trace anomaly~\cite{Christensen:1977jc,Navarro-Salas:1995lmi,Meda:2021zdw,Bousso:1997cg,Balbinot:1999ri}). There has been a renewed interest in the community after the measurement of anomaly signatures in the thermal transport of Weyl semimetals, both for the chiral~\cite{Vu:2021} and mixed chiral-gravitational anomalies~\cite{Gooth:2017mbd}.

In this letter we are particularly interested in the trace anomaly, also known as Weyl or conformal anomaly~\cite{Capper:1974ic,Capper:1975ig}, which arises in the following way: for Weyl-invariant theories, the symmetry forces the corresponding stress tensor to be classically traceless on-shell (see the discussion below). The Weyl anomaly then consists in the emergence of a non-vanishing trace in the quantum theory~\cite{Christensen:1978yd,Fujikawa:1980rc,GamboaSaravi:1983ge,Bastianelli:1991be,Deser:1993yx,Asorey:2003uf}. 

Generically, it is a function of the background fields to which the theory is coupled~\cite{Deser:1976yx,Duff:1977ay}. Such a function is made of terms that must satisfy the usual constraints of the Wess--Zumino consistency conditions~\cite{Wess:1971yu}. Moreover, all possible contributions have been classified and computed using algebraic methods in arbitrary dimensions~\cite{Boulanger:2007ab}. 

Less studied is the fate of the trace anomaly once conformal invariance is broken in a theory.
In contrast to the conformal case, a concrete generalization of the anomaly in this context is subtle: While the trace of the quantized stress tensor is clearly well-defined, the identification of its intrinsically quantum, anomalous part has been debated~\cite{Duff:1993wm, casarin_godazgar_nicolai_plb_2018}. It is the purpose of this letter to clarify this definition.

Before going into the details, it should be emphasized that the understanding of this subject is far-reaching. For example, since in extended supergravity the anomalies of different local and global symmetries are coupled, to obtain a consistent theory the full trace anomaly must vanish. Under this requirement, candidate theories have been built out of the $N = 4$ multiplet of supergravity and an arbitrary number of $N = 4$ Yang-Mills multiplets~\cite{Christensen:1980ee,Henningson:1998gx,Aharony:1999ti}, as well as for $N \geq 5$ Poincar{\'e} supergravities~\cite{Meissner:2016onk,Meissner:2017qwm}, having to assume a certain choice of anomaly coefficients for the not always conformally invariant higher-spin sectors.

A further motivation follows from the guiding principles of the renormalization group, which describes critical systems in terms of conformal theories~\cite{Wilson:1974mb}. Building real systems at criticality is hard, if not impossible. Even if the anomaly could vanish at a fixed point also in the quantum theory of models breaking scale invariance \cite{Morris:2018zgy}, the understanding of theories along the renormalization flow requires the comprehension of non-conformal theories~\cite{Rosten:2018cyr,Bonanno:2020bil}. 

Returning to the generalization of the anomaly, in the literature~\cite{Duff:1993wm, casarin_godazgar_nicolai_plb_2018} the formula
\begin{equation}
\label{eq:definition_Nicolai}
\mathcal{A}_{D} \coloneqq \lim_{n \to 4} \left[ g_{(4)}^{\mu\nu} \expect*\Big{ T_{\mu\nu}(x) } - \expect{ g_{(n)}^{\mu\nu} T_{\mu\nu} } \right]
\end{equation}
has been proposed in the context of dimensional regularization (though it is considered to be more general in the folklore). While Eq.~\eqref{eq:definition_Nicolai} encompasses the general idea that the classical trace, which breaks conformal symmetry, should somehow be subtracted (the second term on the right-hand side), its physical origin is rather obscure. In addition, we find Eq.~\eqref{eq:definition_Nicolai} unsatisfactory for at least two further reasons.

First, it is defined only for a specific renormalization method. If the anomaly possesses a physical meaning, it should be independent of the computational strategy employed; a \emph{sine qua non} condition to check this independence is to have a reliable definition valid for arbitrary schemes. If we choose a non-metric-modifying scheme (essentially, anything else than dimensional regularization), Eq.~\eqref{eq:definition_Nicolai} trivially vanishes. Thus, requiring that the definition should be valid also for Weyl-invariant theories, we are lead to an inconsistency. Related to this critique, it is not even clear how one should extend an arbitrary given metric background to $n$ dimensions, whether this extension is unique or even if different possibilities would lead to the same result.

The second concern has to do with the equation of motion (EOM): it is not specified whether one may use it to simplify any of the terms on the right-hand side before taking the expectation value, an operation which is performed in Ref.~\cite{casarin_godazgar_nicolai_plb_2018}. 
The educated guess, that the EOM should not be used at all, turns out to be wrong, even for classically conformally invariant theories. 

In this letter we will explicitly show that an unambiguous, generalized definition of the trace anomaly that avoids the above-mentioned criticisms is possible for non-conformal models. We will illustrate how this can be achieved in the simple example of a scalar field, its generalization to other spins being straightforward.

\section{(Non-)conformal anomalies in curved spacetime}

Consider a non-minimally coupled, real scalar field of mass $m$ defined on an $n$-dimensional Lorentzian manifold, whose action reads
\begin{equation}
\label{eq:action}
S = - \frac{1}{2} \int \left( \nabla^\mu \phi \nabla_\mu \phi + m^2 \phi^2 + \xi R \phi^2 \right) \sqrt{-g} \total^n x \eqend{.}
\end{equation} 
The geometric quantities in this action are the Ricci scalar $R$, the compatible covariant derivative $\nabla_\mu$ and the determinant $g = \det(g_{\mu\nu})$, all of them  corresponding to the metric $g_{\mu\nu}$. We define as usual the (local) Weyl transformation parameterized by $\sigma(x)$ as
\begin{subequations}
\label{eq:Weyl_transform}
\begin{align}
\phi &\to \phi' \coloneqq \mathe^{w \sigma(x)} \phi \eqend{,} \label{eq:Weyl_transform_field} \\
g_{\mu\nu} &\to g'_{\mu\nu} \coloneqq \mathe^{-2 \sigma(x)} g_{\mu\nu} \eqend{.} \label{eq:Weyl_transform_metric}
\end{align}
\end{subequations}
The parameter $w$ is called the Weyl weight and depends on the type of field being analyzed; for a scalar field it reads $w_\phi = \tfrac{n-2}{2}$. As is well-known, the action in Eq.~\eqref{eq:action} is Weyl invariant only if we choose special values for the mass and the non-minimal coupling, namely $m = 0$ and $\xi = \xi_\text{cc} \coloneqq \tfrac{n-2}{4(n-1)}$; we will refer to them as the conformal couplings, leaving the adjective non-conformal for any other situation.
By virtue of the symmetry, in the Weyl-invariant case the stress tensor is classically traceless on-shell, while for arbitrary couplings it contains terms proportional to $\xi-\xi_\text{cc}$ or $m^2$.

At the quantum level, the fields are promoted to operators, whose expectation values are to be computed. For the general strategy of the computation, we define the composite operators $\Phi^{(2)} \coloneqq \phi^2$ and $\Phi^{(3)}_{\mu\nu} \coloneqq \nabla_\mu \phi \nabla_\nu \phi$, which are the independent operators that appear in the stress tensor of our free theory, and write
\begin{equation}
\label{eq:Tmunu}
\begin{split}
T_{\mu\nu} &\coloneqq \frac{-2}{\sqrt{-g}} \frac{\delta S}{\delta g^{\mu\nu}} \\
&= \left( \delta_\mu^\rho \delta_\nu^\sigma - \frac{g_{\mu\nu}}{2}  g^{\rho\sigma} \right) \Phi^{(3)}_{\rho\sigma} - \frac{g_{\mu\nu}}{2}  m^2 \Phi^{(2)} \\
&\quad- \xi \left( \nabla_\mu \nabla_\nu - g_{\mu\nu} \nabla^2 - R_{\mu\nu} + \frac{1}{2} g_{\mu\nu} R \right) \Phi^{(2)} \eqend{,}
\end{split}
\end{equation}
where $\nabla^2$ denotes the curved spacetime d'Alembertian and $R_{\mu\nu}$ is the Ricci tensor. By the same token, we have the direct, unregularized trace
\begin{equation}
\label{eq:traceTmunu}
\begin{split}
T &= - \frac{n-2}{2} g^{\rho\sigma} \Phi^{(3)}_{\rho\sigma} - \frac{1}{2} n m^2 \Phi^{(2)} \\
&\quad+ \xi (n-1) \nabla^2 \Phi^{(2)} - \xi \frac{n-2}{2} R \Phi^{(2)} \eqend{.}
\end{split}
\end{equation}

Recall that in quantum field theory expectation values are usually divergent and that of the stress tensor is no exception; thus, a renormalization process is required to give them a physical meaning. For the sake of generality, we are not going to fix a scheme until we perform the explicit computation of the anomaly. After subtracting the infinities, ``finite renormalizations'' are still possible~\cite{hollands_wald_cmp_2001}. In effect, even after requiring locality and covariance, the existing freedom in the definition of the operators (Wick products) is larger in curved spacetime than in Minkowski, given that it corresponds to polynomials in the metric, curvature and other invariants of the theory (see the supplemental material for the $n = 4$ case~\cite{supplemental_arxiv}). This freedom can be used to impose further conditions on the renormalized operators; in particular, it is natural to require the conservation of the renormalized stress tensor~\cite{hollands_wald_rmp_2005}. At first sight, it is curious that some coefficients remain undetermined even after imposing conservation; however, it is easy to verify that they give rise to trivial contributions to the anomaly ($m^4$, $m^2R$ and $\nabla^2 R$), i.e., they can be removed by adding local counterterms to the action.

Now let us try to generalize the definition~\eqref{eq:definition_Nicolai} to arbitrary renormalization schemes. To render it consistent, a natural try is to consider the addition of a supplementary contribution, which, taking into account the quantum nature of the anomaly, should vanish at the classical level. Hence, an instinctive guess is that it should involve the EOM, whose product with other operators has an expectation value that does not necessary vanish~\cite{Collins:1984}. Following Ref.~\cite{hollands_wald_cmp_2001}, if we require analyticity in the parameters of the theory and the geometric invariants, dimensionality determines the only available operator, which expressed in terms 
of $\Phi^{(2)}$ and $\Phi^{(3)}$ is
\begin{equation}
E \coloneqq \nabla^2 \Phi^{(2)} - 2 \left( m^2 + \xi R \right) \Phi^{(2)} - 2 g^{\rho\sigma} \Phi^{(3)}_{\rho\sigma} \eqend{.}
\end{equation}

For this reason we add the terms $\beta g_{\mu\nu} E$ and $\beta_\text{tr} E$, respectively, to the stress tensor~\eqref{eq:Tmunu} and its classical trace~\eqref{eq:traceTmunu}. Generalizing the definition in Eq.~\eqref{eq:definition_Nicolai} and renormalizing appropriately, for non-metric-modifying schemes we obtain 
\begin{equation}
\label{eq:anomaly_scalar_field_partially_on_shell}
\begin{split}
&\mathcal{A}_{\beta,\beta_{\rm tr}} \coloneqq g^{\mu\nu} \expect{ (T_{\mu\nu} + \beta g_{\mu\nu} E)^{\text{ren}} } - \expect{ (T + \beta_\text{tr} E)^\text{ren}} \\
&= \frac{n \beta - \beta_\text{tr}}{n-2 + 4 n \beta} \Big[ 4 g^{\mu\nu} \expect{ T_{\mu\nu}^\text{ren} +\beta g_{\mu\nu} E^{\text{ren}}} \\
&\hspace{1em}+ 4 m^2 \expect*\big{ \Phi^{(2),\text{ren}} } + [ (n-2) - 4 (n-1) \xi ] \nabla^2 \expect*\big{ \Phi^{(2),\text{ren}} } \Big] \eqend{,} \raisetag{6.2em}
\end{split}
\end{equation}
where we have expressed $\Phi^{(3),\text{ren}}_{\mu\nu}$ in terms of $(T_{\mu\nu}+\beta g_{\mu\nu} E)^\text{ren}$\footnote{For the scalar field, this can always be done if $n (1 + 4 \beta) \neq 2$, which we assume in the following.}.

It is clear that the so-defined anomaly in general depends on the parameters $\beta$ and $\beta_\text{tr}$, which control the on-shellness of the operators and lead to different results~\footnote{In the non-conformal case, it additionally depends on the renormalization freedom of the operator $\Phi^{(2),\text{ren}}$~\cite{supplemental_arxiv}; however, as mentioned above, the coefficients of these terms are not predictions of the theory.}. We argue in the following that there is a single physically acceptable choice of these coefficients.

To begin with, notice that the role of the $\beta$ inside the brackets in Eq.~\eqref{eq:anomaly_scalar_field_partially_on_shell} is to render the stress tensor conserved if chosen appropriately. Indeed, for a large class of schemes it has been shown that the finite renormalization leading to conservation corresponds to a particular value of $\beta$. For renormalization methods whose counterterms are themselves conserved~\footnote{Among these methods one finds the heat-kernel regularization~\cite{supplemental_arxiv} and dimensional regularization with minimal subtraction, as well as the $\zeta$-function regularization~\cite{moretti_cmp_1999, moretti_jmp_1999}.}, $\beta = - \frac{1}{4}$ yields a conserved stress tensor~\cite{moretti_cmp_1999}. On the other hand, for methods whose counterterms are not conserved such as point splitting and Hadamard subtraction, a different $\beta$ is required~\cite{moretti_cmp_2003}.

In this way, conservation restricts the scheme dependence to the overall factor in Eq.~\eqref{eq:anomaly_scalar_field_partially_on_shell}. The only possibility to have a scheme-independent contribution is thus to choose
\begin{equation}
\beta_\text{tr} = \hat{\beta}_\text{tr} \coloneqq - \frac{n-2}{4} \eqend{,}
\end{equation}
since then the overall factor becomes $\beta$-independent. Importantly, this assertion does not depend on whether we are working with a conformal or a non-conformal theory.

Combining these arguments, we are led to define the generalized anomaly as
\begin{equation}
\label{eq:anomaly_final}
\mathcal{A} \coloneqq g^{\mu\nu} \expect{ T_{\mu\nu}^\text{ren} }^\star - \expect{ T^\text{ren}}^\star - \hat\beta_{\rm tr}  \expect{E^\text{ren}}^\star \eqend{,}
\end{equation}
where the star means that the renormalization employed is such that the resulting stress tensor is conserved. In this expression, $ \hat{\beta}_\text{tr}$ will depend on the field content (see below the result for fermion and vector fields), but we insist that it does not depend on the scheme. In addition, the usual definition of the trace anomaly, i.e., $g^{\mu\nu} \expect{ T_{\mu\nu}^{\text{ren}}}$ in the Weyl-invariant case, is a special case of Eq.~\eqref{eq:anomaly_final}; this can be easily corroborated, given that in this situation $T$ reduces to an expression proportional to $E$, which is exactly cancelled by the $\beta_\text{tr}$ term. It is curious, however, that in the non-conformal case the classical trace is still partially off-shell, 
\begin{equation}
\label{eq:final_trace}
T + \hat{\beta}_\text{tr} E = (n-1) (\xi-\xi_\text{cc}) \nabla^2 \phi^2 - m^2 \phi^2 \eqend{.}
\end{equation}
This fact seems to obscure the physical origin of $\hat\beta_{\rm tr}$, rendering it unintelligible. As we will show shortly, physical cogency can be found within the functional integral approach.

\section{Trace anomaly from the functional integral}

The understanding of trace and axial anomalies in terms of Jacobians of functional integrals was developed by Fujikawa~\cite{Fujikawa:1980vr}. In the case of trace anomalies, the discussion naturally involved classically Weyl--invariant theories. We will show below that the idea can be generalized for non-conformal theories, leading to fixed off-shell extension parameters $\beta$ and, more importantly, $\beta_\text{tr}$.

In the functional integral formalism, we consider the effective action $\Gamma[g^{\mu\nu}]$, defined in a Lorentzian manifold as
\begin{equation}
\label{eq:gamma_def}
\Gamma[g^{\mu\nu}] \coloneqq - \mathi \ln \int \mathe^{\mathi S[\phi, g^{\mu\nu}]} \mathcal{D} \phi \eqend{.}
\end{equation}
The corresponding  quantum stress tensor is defined as in Eq.~\eqref{eq:Tmunu}, replacing $S$ with $\Gamma$. In these definitions, a regulator is obviously implicit and suitable counterterms must be added to the action $S$ to obtain a finite result. Using functional techniques, Fujikawa~\cite{Fujikawa:1980rc, Fujikawa:2004cx} has pointed out that the appropriate measure to obtain a covariant effective action involves a redefinition of the quantum fields, $\hat{\phi} \coloneqq (-g)^\frac{1}{4} \phi$; as shown by Toms~\cite{Toms:1986sh}, this redefinition corresponds to the choice of an orthonormal frame in field space. In this way, the modified effective action
\begin{equation}
\label{eq:hatgamma_def}
\hat{\Gamma}[g^{\mu\nu}] \coloneqq - \mathi \ln \int \mathe^{\mathi S[ (-g)^{-\frac{1}{4}} \hat{\phi}, g^{\mu\nu} ]} \mathcal{D} \hat{\phi}
\end{equation}
automatically leads to a conserved stress tensor, which is related to the previous one as follows:
\begin{equation}
\expect*\big{ \hat{T}_{\mu\nu} } \coloneqq - \frac{2}{\sqrt{-g}} \frac{\delta \hat{\Gamma}}{\delta g^{\mu\nu}} = \expect{ T_{\mu\nu} } - \frac{g_{\mu\nu}}{2 \sqrt{-g}} \expect*\Big{ \phi \frac{\delta}{\delta \phi} S } \eqend{.}
\end{equation}
Since $\phi \frac{\delta}{\delta \phi} S = \frac{1}{2} \sqrt{-g} E$, the conservation of the stress tensor is thus equivalent to an off-shell extension of the stress tensor with $\beta = - \frac{1}{4}$, as we have anticipated in the previous section (by defining the stress tensor as the variation of an action, the included counterterms are covariantly conserved).

If we change the variables in the functional integral~\eqref{eq:hatgamma_def} according to the Weyl transformation of the fields~\eqref{eq:Weyl_transform}, we obtain
\begin{equation}
\hat{\Gamma}[\mathe^{2\sigma} g^{\mu\nu}] = - \mathi \ln \int \mathe^{\mathi S[ \mathe^{w_\phi \sigma} (-g)^{-\frac{1}{4}} \hat{\phi}, \mathe^{2\sigma} g^{\mu\nu} ]} \hat{\mathcal{J}}[\sigma] \mathcal{D} \hat{\phi} \eqend{,}
\end{equation}
where we have taken into account that a Jacobian $\hat{\mathcal{J}}$ may arise in the measure. Evidently, for $\sigma = 0$, $\hat{\mathcal{J}}$ should be trivial. In order to characterize the system under such a transformation, it is natural to analyze the variations with respect to $\sigma$. This provides a modified Ward--Takahashi identity
\begin{equation}
\label{eq:variation_fujikawa1}
\begin{split}
g^{\mu\nu} \expect*\big{ \hat{T}_{\mu\nu} } &= \frac{\mathi}{\sqrt{-g}} \left. \frac{\delta}{\delta \sigma(x)} \hat{\mathcal{J}}[\sigma] \right\rvert_{\sigma = 0} \\
&\quad- \frac{w_\phi}{\sqrt{-g}} \expect*\Big{ \phi \frac{\delta}{\delta \phi} S } + g^{\mu\nu} \expect{ T_{\mu\nu} } \eqend{.}
\end{split}
\end{equation}

Undoubtedly, the contribution from the Jacobian is a purely quantum contribution; moreover, it is the one responsible for the trace anomaly for Weyl-invariant theories. We therefore define
\begin{equation}
\begin{split}
\label{eq:anomaly_our_definition}
\mathcal{A}_\text{F}(x) &\coloneqq \frac{\mathi}{\sqrt{-g}} \left. \frac{\delta}{\delta \sigma(x)} \hat{\mathcal{J}}[\sigma] \right\rvert_{\sigma = 0} \\
&= \frac{1}{\sqrt{-g}} \left( w_\phi - \frac{n}{2} \right) \expect*\Big{ \phi \frac{\delta}{\delta \phi} S } = - \frac{1}{2} \expect{ E } \eqend{,}
\end{split}
\end{equation}
where the last line is a consequence of Eq.~\eqref{eq:variation_fujikawa1}. This formula is all we need in order to determine the coefficient $\beta_\text{tr}$ in the general formula~\eqref{eq:anomaly_scalar_field_partially_on_shell}. In fact, if in Eq.~\eqref{eq:anomaly_scalar_field_partially_on_shell} we express $g^{\mu\nu} \expect{ T_{\mu\nu}^\text{ren} }$ in terms of $\expect{ E^\text{ren} }$ and $\expect{ \Phi^{(2),\text{ren}} }$, we obtain $\mathcal{A}_{\beta,\beta_{\text{tr}}} = ( n \beta - \beta_\text{tr} ) \expect{ E^\text{ren} }$. Since we have already determined that $\beta = - \frac{1}{4}$, it follows that $\beta_\text{tr} = \hat{\beta}_\text{tr}$. Hence, once we properly identify the anomaly as arising from the Jacobian {\`a} la Fujikawa, the degree of on-shellness is not arbitrary, but fixed even for non-conformal theories.

\section{Conclusions}

We have shown that the frequently acknowledged generalization of the trace anomaly for non-conformal theories, Eq.~\eqref{eq:definition_Nicolai}, entails some intricacies that are often overlooked. We have shown that the \emph{unique} scheme-independent possible definition of the anomaly is Eq.~\eqref{eq:anomaly_final}. In particular, it is not tied to the use of dimensional regularization, which is not always a suitable tool. Conceptually, the physical origin of the anomaly is (once more) the Jacobian arising in the measure after a Weyl transformation of the variables.

Notice that the last term in Eq.~\eqref{eq:anomaly_final} can be understood as a special off-shell extension of the classical trace of the stress tensor. In general, this term is relevant and must be included, even though it might vanish in some regularization and renormalization schemes. Nevertheless, in some cases it cancels out with the second term in Eq.~\eqref{eq:anomaly_final}, as can be readily seen from the expression~\eqref{eq:final_trace} for the scalar field, resulting in the formula associated to the anomaly for classically Weyl-invariant theories.

For a scalar field in $n = 4$, using our definition~\eqref{eq:anomaly_final} and the heat-kernel renormalization technique
described in detail in the supplemental material~\cite{supplemental_arxiv} (see also references~\cite{Freedman:2011hp, DeWitt:2003,hollands_wald_cmp_2002, Christensen:1976vb} therein), we find the anomaly 
\begin{equation}
\label{eq:our_explicit_anomaly}
\mathcal{A}^{n=4}_{s=0} = \frac{3 C^{\mu\nu\rho\sigma} C_{\mu\nu\rho\sigma} - \mathcal{E}_4 + 5 (1-6\xi)^2 R^2}{360 (4 \pi)^2} \eqend{,}
\end{equation}
which is written in terms of the square of the four-dimensional Weyl tensor $C$ and the Euler density: 
\begin{align}
C^{\mu\nu\rho\sigma} C_{\mu\nu\rho\sigma} =& R^{\mu\nu\rho\sigma} R_{\mu\nu\rho\sigma} - 2 R^{\mu\nu} R_{\mu\nu} + \frac{1}{3} R^2 \eqend{,} \\
\mathcal{E}_{4} \coloneqq& R^{\mu\nu\rho\sigma} R_{\mu\nu\rho\sigma} - 4 R^{\mu\nu} R_{\mu\nu} + R^2 \eqend{.}
\end{align}
Note that in the expression~\eqref{eq:our_explicit_anomaly} we have dismissed all the trivial contributions to the anomaly. The result is thus mass independent, which could have been predicted from the available terms that can be built with the right dimensions using geometric invariants. Moreover, it is consistent with the massless result in Ref.~\cite{casarin_godazgar_nicolai_plb_2018}, which was obtained using dimensional regularization and a perturbative expansion around flat space, together with a dose of intuition to use the right amount of on-shellness. Notice also the appearance of the $R^2$, which, contrary to the  Weyl and Euler density terms, signals the violation of the Wess--Zumino consistency condition~\cite{Bonora:1983ff}. This does not imply at all an inconsistency in our result, given that the Wess--Zumino consistency cannot be recast as a second order variation (using our definition). 

Importantly, the arguments that we have presented for a scalar field can be straightforwardly extended to higher spins as well, except for the fact that the choice of the coefficients may differ. For massive Dirac fermions (Ref.~\cite{Gorbar:2003yt} has been pointed out to us, which provides relevant computations for massive fermions and vectors), the expression for the anomaly is encoded in Eq.~\eqref{eq:anomaly_final}, by simply replacing the symbols with $E^\psi = \frac{1}{2 \sqrt{-g}} \left( \frac{\delta_\text{R} S}{\delta \psi} \psi + \bar{\psi} \frac{\delta_\text{L} S}{\delta \bar\psi} \right)$ and $\hat\beta^{s=1/2}_\text{tr} = 1 - n$; $\beta^{s=1/2} = -1$ ensures the conservation of the stress tensor. This coincides with the expression given by Fujikawa~\cite{Fujikawa:1980rc} and, after removing the trivial terms, a direct computation shows that it coincides with the trace anomaly for massless Dirac fermions which is found in the literature. 

As a further example, consider a free massive, non-minimally coupled Abelian vector field $A_\mu$. A rather general action for this field reads
\begin{equation}
\label{eq:action_vector}
\begin{split}
S_{s=1} &= - \frac{1}{2} \int \bigg[ \frac{1}{2} F_{\mu\nu} F^{\mu\nu} + ( m^2 + \xi R ) A^\mu A_\mu \\
&\quad+ (\zeta-1) A_\mu A_\nu R^{\mu\nu} + \alpha^{-1} \left( \nabla_\mu A^\mu \right)^2 \bigg] \sqrt{-g} \total^n x \eqend{,}
\end{split} \raisetag{4.8em}
\end{equation}
where $\alpha$, $\zeta$ and $\xi$ are arbitrary coefficients parametrizing the different non-minimal and gauge fixing terms, $m$ is the mass and as usual we have defined the field strength $F_{\mu\nu} \coloneqq \nabla_\mu A_\nu - \nabla_\nu A_\mu$. From this action one can derive the corresponding EOM
\begin{equation}
\label{eq:EOM_vector}
\begin{split}
0 = M_{\mu} &\coloneqq \left(  \nabla^2 - m^2 - \xi R \right) A_\mu
\\
&\quad- \left[ \zeta R_{\mu\nu} + (1-\alpha^{-1}) \nabla_\mu \nabla_\nu \right] A^\nu \eqend{,}
\end{split}
\end{equation}
as well as the stress tensor, whose explicit expression can be found in the supplemental material~\cite{supplemental_axiv}. The vectorial character of the field implies that we have two different ways of implementing the EOM~\footnote{Once more, we are following the natural hypothesis in Ref.~\cite{hollands_wald_cmp_2001}, i.e. we are considering only the addition of terms which are polynomials in the physical quantities.}, 
which we parametrize with coefficients $\alpha_{i=1,2}$:
\begin{equation}
T_{\mu\nu}^{\alpha_i} \coloneqq T_{\mu\nu} + \alpha_1 \left( A_\mu M_\nu + A_\nu M_\mu \right) + 2 \alpha_2 g_{\mu\nu} A_\rho M^\rho \eqend{.}
\end{equation}

Following the above-described procedure and fixing for simplicity  $n=4$ and $\alpha=1$ (Feynman--'t Hooft gauge), using heat-kernel renormalization we find $\alpha_1 = 1/2$, $\alpha_2 = - 1/4$, $\hat\beta^{s=1}_{\text{tr}} = 0$ and the following anomaly (we are including the ghost contribution present in the gauged case as discussed in the supplemental material~\cite{supplemental_arxiv}):
\begin{widetext}
\begin{equation}
\mathcal{A}^{n=4}_{s=1} = \frac{1}{360 (4 \pi)^2} \Big[ (90 \zeta^2-54) C^{\mu\nu\rho\sigma} C_{\mu\nu\rho\sigma} 
- (90\zeta^2-28) \mathcal{E}_4 + 60 \bigl( 12 \xi^2 + (\zeta-1) \zeta + \xi (6 \zeta-4) \bigr) R^2\Big] \eqend{.}
\end{equation}
\end{widetext}

An adaptation of these results sheds light on present discussions regarding the trace anomaly for massless Weyl fermions. In effect, the appropriate definition of the anomaly has recently become the kernel of the debate~\cite{Coriano:2023cvf,Abdallah:2023cdw,Bonora:2022izj,Abdallah:2021eii}; this might have tremendous consequences on the unitarity of some theories (for instance, the standard model with at least one massless neutrino), since it implies the occurrence of an imaginary term in the anomaly. Employing our definition for the anomaly, we see that there is indeed a term corresponding to the expectation value of the trace $\expect{T^\mu{}_\mu}$, as claimed in Ref.~\cite{Bonora:2022izj}. However, one should not forget to also include the term proportional to the EOM, which exactly counterbalances it and precludes the appearance of imaginary terms.

We can also extend our discussion to other dimensions, including the subtler $n=2$ case. 
Following the same lines as before, for a number $N_0$ of scalars and $N_{1/2}$ of fermions we obtain
\begin{equation}
\mathcal{A}^{n=2} = \frac{(1-6\xi) N_0 + N_{1/2}}{24 \pi} R.
\end{equation}
A deeper discussion of this outcome is left to the supplemental material~\cite{supplemental_arxiv} (see also references~\cite{Barvinsky:1990up, Avramidi:1990je, Ribeiro:2018pyo, Polyakov:1981rd} therein).

In view of our results, it is not surprising that some ambiguities are present in the literature of trace anomaly cancellations in supergravity~\cite{Meissner:2017qwm}. The fact is that several of these computations have been done using the $b_{n}$ coefficient of the heat kernel, which can be related to the quantum trace of the stress tensor, while the theory involves supermultiplets whose higher-spin components possess non-Weyl-invariant actions~\cite{Fradkin:1982bd}. Hence, according to Eq.~\eqref{eq:anomaly_final}, those computations are in general not universal and might thus be plagued by ambiguities; Understanding the exact mechanism behind them is not straightforward and work has still to be done.

The last question we want to tackle is what happens for interacting theories.
The application of the Fujikawa method to interacting theories is involved, since the recasting of the functional integral measure in terms of the coefficients accompanying the basis of eigenfunctions would imply solving nonlinear differential equations. In spite of that, without computing the explicit Jacobian, one can consider expressions in terms of composite operators, i.e., analogous to Eq.~\eqref{eq:anomaly_scalar_field_partially_on_shell}. These are tractable in a perturbative computation at the multiloop level and are thus expected to be compatible with an effective field theory approach, as long as one is able to introduce order by order the appropriate terms proportional to the EOM. An understanding of the generalized trace anomaly for non-conformal theories would provide a new avenue to $c$-theorems, given that the existing proof in $n=4$~\cite{Komargodski:2011vj} relies on the matching of the trace anomaly between appropriate infrared and ultraviolet conformal field theories;
instead, our anomaly could provide a connection throughout the whole flow without the need to introduce spurious interactions.

\textit{Note added.}--- 
Regarding the (odd) trace anomaly for Weyl fermions, a new article has appeared that support our claims~\cite{Larue:2023tmu}. The interested reader might also compare it with the Pauli--Villars computation~\cite{Bastianelli:2019zrq}, which leads to the same conclusion.

\begin{acknowledgments}
The authors are grateful to F.~Bastianelli, L.~Casarin, L. Dixon, D.~Mazzitelli, R.~Schützhold and O.~Zanusso for inspiring discussions.
SAF acknowledges the support from Helmholtz-Zentrum Dresden-Rossendorf, PIP 11220200101426CO Consejo Nacional de Investigaciones Científicas y Técnicas (CONICET), Project 11/X748 (UNLP) and Deutsche Forschungsgemeinschaft (DFG, German Research Foundation) --- Project-ID 278162697 within the SFB 1242. 
MBF acknowledges funding by the Deutsche Forschungsgemeinschaft (DFG, German Research Foundation) --- project no. 396692871 within the Emmy Noether grant CA1850/1-1.
\end{acknowledgments}

\appendix

\section{SUPPLEMENTAL MATERIAL}
We follow the conventions of Ref.~\cite{Freedman:2011hp}, which corresponds to the so-called ``+++'' convention of Misner, Thorne and Wheeler. We use Planck units in which $\hbar = 1 = c$ and work in a Lorentzian spacetime of dimension $n$.

\section{Freedom in the composite operators}

As an example, in $n=4$, the ambiguities for the composite operators at hand read
\begin{subequations}
\label{eq:phi23_renormalization_freedom}
\begin{align}
\Phi^{(2),\text{ren}} &\to \Phi^{(2),\text{ren}} + c_{0,0} m^2 + c_{2,0} R \eqend{,} \label{eq:phi2_renormalization_freedom} \\
\begin{split}
\Phi^{(3),\text{ren}}_{\mu\nu} &\to \Phi^{(3),\text{ren}}_{\mu\nu} + \sum_{i=1}^8 c_{4,i} C^{(4,i)}_{\mu\nu} \\
&\qquad + m^2  \sum_{i=1}^2 c_{2,i} C^{(2,i)}_{\mu\nu} + c_{0,1} m^4 g_{\mu\nu} \eqend{,} \label{eq:phi3_renormalization_freedom}
\end{split}
\end{align}
\end{subequations}
where the $c_{k,i}$ are numerical constants that may depend on the dimensionless coupling parameter $\xi$, while the $C^{(d,i)}_{\mu\nu}$ are combinations of curvature tensors and their derivatives of engineering dimension $d$.

In Eq.~\eqref{eq:phi23_renormalization_freedom}, the freedom in the composite operators has been parameterized in terms of the following quantities:
\begin{equation}
\begin{alignedat}{2}
C^{(2,1)}_{\mu\nu} &\coloneqq m^2 R_{\mu\nu} \eqend{,}
&\quad C^{(2,2)}_{\mu\nu} &\coloneqq g_{\mu\nu} m^2 R \eqend{,} \\
C^{(4,1)}_{\mu\nu} &\coloneqq g_{\mu\nu} R_{\alpha\beta\gamma\delta} R^{\alpha\beta\gamma\delta} \eqend{,}
&\quad C^{(4,2)}_{\mu\nu} &\coloneqq g_{\mu\nu} R_{\alpha\beta} R^{\alpha\beta} \eqend{,} \\
C^{(4,3)}_{\mu\nu} &\coloneqq g_{\mu\nu} R^2 \eqend{,}
&\quad C^{(4,4)}_{\mu\nu} &\coloneqq g_{\mu\nu} \nabla^2 R \eqend{,} \\
C^{(4,5)}_{\mu\nu} &\coloneqq R^{\alpha\beta} R_{\alpha\mu\beta\nu} \eqend{,}
&\quad C^{(4,6)}_{\mu\nu} &\coloneqq R_{\mu\nu} R \eqend{,} \\
C^{(4,7)}_{\mu\nu} &\coloneqq \nabla^2 R_{\mu\nu} \eqend{,}
&\quad C^{(4,8)}_{\mu\nu} &\coloneqq \nabla_\mu \nabla_\nu R \eqend{.}
\end{alignedat}
\end{equation}

\section{Heat-kernel regularization}
\label{app:HK}

In the heat-kernel regularization~\cite{DeWitt:2003}, one writes the Feynman propagator as a modified Laplace transform of the heat kernel $K(x,x';\tau)$:
\begin{equation}
\label{eq:heat_kernel_feynman}
G^\text{F}(x,x') = - \mathi \lim_{\epsilon \to 0^+} \int_0^\infty \tau^\alpha K(x,x';\tau) \, \mathe^{- \mathi (m^2 - \mathi \epsilon) \tau} \total \tau \eqend{.}
\end{equation}
The $\epsilon$ parameter enforces the Feynman prescription for the propagator, while $\alpha > 0$ is a regulator whose effect is best appreciated at the coincident limit of the propagator (we will be interested in the $\alpha \to 0^+$ limit).

It is noteworthy to remark that the heat-kernel regularization is a locally covariant method, a category that embraces all methods ensuring that the renormalized quantum fields transform properly under diffeomorphisms~\cite{hollands_wald_cmp_2001,hollands_wald_cmp_2002}. Other examples sharing this property include point splitting and Hadamard subtraction, as well as $\zeta$-function regularization~\cite{moretti_cmp_1999}.

The relation in Eq.~\eqref{eq:heat_kernel_feynman} implies that the expectation values of the composite operators which are quadratic in the fields can be recast in terms of the coincidence limit of the heat kernel~\footnote{Given that we will be considering the coincidence limit, we can simply  employ the Feynman propagator instead of the Wightman function.}. Since some of these operators involve derivatives acting on two different fields, we will have to consider the biscalar character of the heat kernel and only afterwards take the coinciding limit $x'\to x$. In this limit, there will appear some divergences in the composite operators, which will be  renormalized  performing a minimal subtraction of the poles in $\alpha$, an operation that can be done effectively by employing the well-known expansion of the heat kernel.

To illustrate these operations, let us analyze the non-minimally coupled scalar field considered in the body of the article, in which case the heat kernel satisfies the following heat equation with proper time $\tau$~\footnote{In this equation all the covariant derivatives are with respect to $x$;
derivatives with respect to $x'$ will later be denoted by primed indices.}:
\begin{equation}
\partial_\tau K(x,x';\tau) = \mathi \left( \nabla^2 - \xi R \right) K(x,x';\tau) \eqend{.}
\end{equation}
This should be compared with the equation of motion of the field [the mass has been absorbed in the definition~\eqref{eq:heat_kernel_feynman}]:
\begin{equation}
\frac{1}{\sqrt{-g}}\frac{\delta S}{\delta \phi} = \left( \nabla^2 - m^2 - \xi R \right) \phi = 0 \eqend{.}
\end{equation}
Additionally, the heat kernel fulfills a delta-type initial condition,
\begin{align}
\lim_{\tau \to 0^+} K(x,x';\tau) = \frac{\delta^n(x-x')}{\sqrt{-g}} \eqend{.}
\end{align} 

For the heat kernel, we use the well-known Gilkey--Schwinger--DeWitt (GSDW) ansatz, also called Hadamard--Minakshisundaram--DeWitt, which corresponds to an asymptotic expansion for small $\tau$ (or large masses):
\begin{equation}
\label{eq:hk_expansion}
\begin{split}
&K(x,x';\tau) \sim \frac{\sqrt{\Delta(x,x')}}{(4 \pi \tau)^\frac{n}{2}}
\mathe^{\frac{\mathi \sigma(x,x')}{2 \tau}- \mathi \frac{n-2}{4} \pi}\sum_{k=0}^\infty (\mathi \tau)^k A_k(x,x') \eqend{.}
\end{split}
\end{equation}
In this expression, $\Delta$ is the van-Vleck--Morette determinant  and $\sigma$ is Synge's world function, which is proportional to the square of the geodesic distance. Under rather general assumptions, the expansion in Eq.~\eqref{eq:hk_expansion} is locally valid, which is sufficient for our purposes. Notice also that the GSDW coefficients $A_k(x,x')$ inherit the symmetry of the heat kernel in its spatial arguments. 

Replacing the expansion~\eqref{eq:hk_expansion} in the heat kernel equation and comparing quantities at each order in $\tau$, we obtain a recursion relation for the GSDW coefficients outside the diagonal:
\begin{equation}
\label{eq:recursive_relation}
\begin{split}
&\left( \nabla^\nu \sigma \nabla_\nu + k \right) A_k(x,x') = \Bigg[ \nabla^2 - \xi R + \left(\nabla^\nu \ln \Delta\right) \nabla_\nu \\
&\qquad+ \frac{1}{4} \left(\nabla^\nu \ln \Delta\right)^2 + \frac{1}{2} \nabla^2 \ln \Delta \Bigg] A_{k-1}(x,x') \eqend{.}
\end{split}
\end{equation}
Using that $A_0(x,x') = 1$ as a consequence of the initial condition, a solution for the GSDW coefficients in the diagonal can be found recursively~\cite{Christensen:1976vb}. The general theory states that the coincidence limits of these coefficients are local polynomials in the geometric invariants; those relevant to the renormalization of the scalar theory in $n = 4$ dimensions are
\begin{subequations}
\begin{align}
A_0(x,x) &= 1 \eqend{,} \\
A_1(x,x) &= \left( \frac{1}{6} - \xi \right) R \eqend{,} \\
\begin{split}
A_2(x,x) &= \frac{1}{180} \Bigg[ R^{\alpha\beta\gamma\delta} R_{\alpha\beta\gamma\delta} - R^{\alpha\beta} R_{\alpha\beta} \\
&\qquad+ \frac{5}{2} (1-6\xi)^2 R^2 + 6 (1-5\xi) \nabla^2 R \Bigg] \eqend{.}
\end{split}
\end{align}
\end{subequations}
\newpage

As mentioned before, in our computation we will have to deal with the coincidence limit of derivatives of biscalars~\cite{Christensen:1976vb}. One can trade them for derivatives of the coincidence limit of biscalars; this can be done using the recursive relation~\eqref{eq:recursive_relation}, the coincidence limit of the van-Vleck--Morette determinant and Synge's world function~\cite{Christensen:1976vb}, as well as the following properties, which are valid for a symmetric biscalar $f(x,x')$:
\begin{subequations}
\begin{align}
[ g_\mu{}^{\mu'} \nabla_{\mu'} f ] &= \nabla_\mu [ f ]- [ \nabla_\mu f ] \eqend{,} \\
[ \nabla_\mu f ] &= \frac{1}{2} \nabla_\mu [ f ] \eqend{,} \\
\begin{split}
[ \nabla_\nu \nabla^2 f ]&= \nabla^\mu [ \nabla_\mu \nabla_\nu f ] - \frac{1}{4} \nabla_\nu \nabla^2 [ f ] \\
&\qquad+ \frac{1}{2} \nabla_\nu [ \nabla^2 f ] - \frac{1}{2} R_{\nu\mu} \nabla^\mu [ f ] \eqend{,}
\end{split} \\
[ g_\nu{}^{\nu'} \nabla_{\nu'} \nabla_\mu f ] &= \frac{1}{2}\nabla_\mu \nabla_\nu [ f ]- [ \nabla_\mu \nabla_\nu f ] \eqend{.}
\end{align}
\end{subequations}
In these equations, $g_\nu{}^{\nu'}$ is the bivector of parallel displacement and the coincidence limit is compactly written as customarily:
\begin{equation}
[f(x,x')] \coloneqq \lim_{x'\to x} f(x,x') \eqend{.}
\end{equation}

\section{Vector field}

The stress tensor derived from the action~\eqref{eq:action_vector} is
\begin{widetext}
\begin{equation}
\begin{split}
T_{\mu\nu}^{s=1} &= \frac{1}{2} \Big[2 m^2 A_{\mu } A_{\nu }  + 2 \xi A_{\alpha } A^{\alpha } R_{\mu \nu } + 4 (-1 + \zeta) A^{\alpha } A_{(\mu } R_{\nu) \alpha } + 2 \xi A_{\mu } A_{\nu } R + 2 (-1 + \zeta) A_{(\nu } \nabla^2 A_{\mu )} \\
&\qquad- 2 (-1 + \zeta) A_{(\mu } \nabla^{\alpha }\nabla_{\nu) }A_{\alpha } - 2 (-1 + \zeta) A^{\alpha } \nabla_{\alpha }\nabla_{(\mu }A_{\nu) } + 2 \zeta \nabla_{\alpha }A_{\nu } \nabla^{\alpha }A_{\mu } - 2 (1 + \zeta) \nabla^{\alpha }A_{(\nu } \nabla_{\mu) }A_{\alpha } \\
&\qquad- 2 (-1 + \zeta) \nabla_{\alpha }A^{\alpha } \nabla_{(\mu }A_{\nu) } - 4 \alpha^{-1}{} A_{(\nu } \nabla_{\mu )}\nabla_{\alpha }A^{\alpha } - 4 \xi A^{\alpha } \nabla_{(\mu }\nabla_{\nu )}A_{\alpha } + 2 (1 - 2 \xi) \nabla_{\mu }A^{\alpha } \nabla_{\nu }A_{\alpha } \Big] \\
&\quad+ \frac{1}{2} g_{\mu \nu } \Big[- m^2 A_{\alpha } A^{\alpha } -  (-1 + \zeta) A^{\alpha } A^{\beta } R_{\alpha \beta } - \xi A_{\alpha } A^{\alpha } R + (-1 + 2 \alpha^{-1}{} + \zeta) A^{\alpha } \nabla_{\alpha }\nabla_{\beta }A^{\beta } + \zeta \nabla_{\alpha }A_{\beta } \nabla^{\beta }A^{\alpha } \\
&\qquad\qquad+ (-1 + \alpha^{-1}{} + \zeta) \nabla_{\alpha }A^{\alpha } \nabla_{\beta }A^{\beta } + (-1 + \zeta) A^{\alpha } \nabla_{\beta }\nabla_{\alpha }A^{\beta } + 4 \xi A^{\alpha } \nabla^2 A_{\alpha } + (-1 + 4 \xi) \nabla_{\beta }A_{\alpha } \nabla^{\beta }A^{\alpha } \Big] \eqend{,} 
\end{split}   
\end{equation}
\end{widetext}
where we have used an idempotent symmetrization in the indices, $A_{(\mu \nu)} \coloneqq \frac{1}{2} \left( A_{\mu\nu} + A_{\nu\mu} \right)$.

For the vector field, the direct contribution to the anomaly, that is without subtracting the ghost contribution, reads 
\begin{widetext}
\begin{align}
\mathcal{A}^{n=4}_{\text{w/o ghost}} &= \frac{1}{180 (4 \pi)^2}  \left[(13 - 45 \zeta^2) \mathcal{E}_4 + 5 \bigl( 1 + 72 \xi^2 - 6 \zeta + 6 \zeta^2 + 12 \xi (-2 + 3 \zeta) \bigr) R^2 + (-24 + 45 \zeta^2) C^{\mu\nu\rho\sigma} C_{\mu\nu\rho\sigma} \right] \eqend{.}
\end{align}
\end{widetext}
On the other hand, the ghost contribution can be simply written in terms of the anomaly for minimally coupled ($\xi = 0$) scalar fields:
\begin{equation}
\mathcal{A}^{n=4}_{\text{ghost}} = \frac{1}{180 (4 \pi)^2} \left[ - \mathcal{E}_4 + 5 R^2 + 3 C^{\mu\nu\rho\sigma} C_{\mu\nu\rho\sigma} \right] \eqend{.}
\end{equation}

\section{On the two-dimensional results for a scalar field}

As discussed in the main body of the article and in spite of the well-known peculiarities of scalar fields in two dimensions, our results still hold. In effect, the first obstacle that is found is that the finite renormalizations described in Ref.~\cite{hollands_wald_rmp_2005}
are not sufficient to obtain a conserved stress tensor. 
Instead, one should interpret this process as a redefinition of the quantum stress tensor by including a term proportional to the equations of motion~\cite{moretti_cmp_2003}.
Taking this into account, it is straightforward to obtain its renormalized expectation value
\begin{widetext}
\begin{align}
\label{eq:EM_tensor_2d}
\begin{split}
\expect{T_{\mu\nu}^{s=0,n=2}}^\star &= \frac{m^2 g_{\mu \nu } \bigl(2\gamma + 4 \ln(m) \bigr)}{16 \pi} + \frac{(1 - 10 \xi + 30 \xi^2)}{120 m^2 \pi} \left( - g_{\mu\nu} \nabla^2 R+ \nabla_\nu \nabla_\mu R - \frac{g_{\mu\nu} R^2}{4} \right) + \cdots \eqend{,}
\end{split}
\end{align}
\end{widetext}
where the dots indicate contributions with at least a power $m^4$ in the denominator and we have used the existing relation between geometric invariants in two dimensions to express them in terms of the Ricci scalar.

It is important to emphasize the inverse-mass-expansion nature of this expression, which evidently precludes us from taking the massless limit in our simple GSDW approach. This technical issue could be solved by adapting more sophisticated expansions of the heat kernel, such as the covariant derivative expansion~\cite{Barvinsky:1990up}. Unfortunately, a direct application of this technique at the level of the effective action leads to
\begin{equation}
\label{eq:zanusso}
\Gamma^\text{past}_\text{1-loop} = - \frac{1}{96\pi} \int R \frac{1-12\xi+12\xi^2\log \left(\frac{-\nabla^2}{m^2}\right)}{\nabla^2}R \sqrt{-g} \total^2 x \eqend{,}
\end{equation}
where the logarithmic term of Eq.~\eqref{eq:zanusso} evidences an infrared running and the breakdown of the resummation~\cite{Barvinsky:1990up}.

In any case, one can use the stress tensor in Eq.~\eqref{eq:EM_tensor_2d} to obtain the effective action of a massive field by integrating its trace. 
Note then that the first term can be removed by a finite renormalization of the cosmological constant.
As for the second term, it could be tempting to identify it with the first contribution in the large mass expansion of a modified Polyakov action; however, we have no control on the remaining expansion in the mass and it is known that the functional structure of the underlying  effective action is actually more involved~\cite{Avramidi:1990je,Ribeiro:2018pyo}.

These drawbacks disappear once we apply our formula for the anomaly: a highly non-trivial cancellation occurs whose outcome is a mass-independent quantity. Although one could follow the standard procedures to integrate the anomaly~\cite{Polyakov:1981rd},
\begin{equation}
\label{eq:our_polyakov}
I^{\mathcal{A}} = - \frac{1}{96\pi} \int R \frac{1-6\xi}{\nabla^2} R \sqrt{-g} \total^2 x \eqend{,}
\end{equation}
one should keep in mind that in the non-conformal case this is not enough to obtain the full effective action.

\bibliography{letter_v1}

\end{document}